\documentclass[aps,prl,twocolumn,superscriptaddress,
               amsmath,a4paper,floatfix]{revtex4}

\usepackage{graphicx}
\usepackage{dcolumn}
\usepackage{bm}
\usepackage{dcolumn}
\usepackage{subfigure}

\graphicspath{{../}}

\begin{document}

\title{\textbf{
Induced Random Fields in the LiHo$_x$Y$_{1-x}$F$_4$ 
Quantum Ising Magnet in a Transverse Magnetic Field
}}

\author{S.M.A. Tabei}
\affiliation{
Department of Physics and Astronomy, University of Waterloo, Ontario, N2L 3G1, Canada}

\author{M.J.P. Gingras}
\affiliation{
Department of Physics and Astronomy, University of Waterloo, Ontario, N2L 3G1, Canada}
\affiliation{
Department of Physics and Astronomy, University of Canterbury,
Private Bag 4800, Christchurch, New Zealand}

\author{Y.-J. Kao}
\affiliation{
Department of Physics and Astronomy, University of Waterloo, Ontario, N2L 3G1, Canada}
\affiliation{Department of Physics and Center for Theoretical Sciences,
 National Taiwan University, Taipei 10617, Taiwan}

\author{P. Stasiak}
\affiliation{
Department of Physics and Astronomy, University of Waterloo, Ontario, N2L 3G1, Canada}

\author{J.-Y. Fortin}

\affiliation{
Department of Physics and Astronomy, University of Waterloo, Ontario, N2L 3G1, Canada}
\affiliation{Laboratoire de Physique Th\'eorique, Universit\'e
 Louis Pasteur, 67084 Strasbourg, Cedex, France}

\date{\today}

\begin{abstract}
The LiHo$_x$Y$_{1-x}$F$_4$ magnetic material in a transverse magnetic
field $B_{x}\hat x$ perpendicular to the Ising spin direction has long been used
to study tunable quantum phase transitions in a random disordered system.
We show that the $B_{x}-$induced magnetization along the $\hat x$ direction,
combined with the local random dilution-induced 
destruction of crystalline symmetries,
generates, via the predominant dipolar interactions between Ho$^{3+}$ ions,
{\it random fields} along the Ising $\hat z$ direction. 
This identifies  LiHo$_x$Y$_{1-x}$F$_4$ in $B_x$ as a new 
random field Ising system. 
The random fields explain the rapid decrease of the critical temperature in
the diluted ferromagnetic regime and  the smearing
of the nonlinear susceptibility at the spin glass transition 
with increasing $B_{x}$, and
render the $B_{x}-$induced quantum
criticality in  LiHo$_x$Y$_{1-x}$F$_4$ likely inaccessible.

\end{abstract}

\maketitle

Quantum phase transitions (QPTs) occur near absolute zero temperature
and are driven by quantum mechanical fluctuations associated with
the Heisenberg uncertainty principle and
not by thermal fluctuations as in classical phase transitions~\cite{Sachdev,Sondhi}.
The transverse field Ising model (TFIM)~\cite{deGennes,Pfeuty} with Hamiltonian
${{\cal H}_{\rm TFIM}=-\frac{1}{2}\sum_{(i,j)}
J_{ij}\, \sigma_i^z  \sigma_j^z - \Gamma \sum_{i} \sigma_i^x}$,
where $\sigma_i^\mu$ ($\mu=x,y,z$) are Pauli matrices, is 
the simplest theoretical model that exhibits a QPT~\cite{Sachdev,Pfeuty,Sen}.
The field $\Gamma$ transverse to the Ising $\hat z$ direction
causes quantum tunneling between the spin-up and spin-down eigenstates of
$\sigma_i^z$. These spin fluctuations decrease the critical temperature $T_c$ 
at which the spins develop either conventional long-range order
or, for random ferromagnetic and antiferromgnetic $J_{ij}$,
 a spin glass (SG) state with randomly frozen
spins below a $T_g$.
At a critical field $\Gamma_c$, $T_c$ or $T_g$
 vanishes, and a 
quantum  phase transition between either
a long-range ordered or SG state and a 
quantum paramagnet (PM) insues.

The phenomenology of both the disorder-free
and random TFIM has been extensively investigated 
in the LiHo$_x$Y$_{1-x}$F$_4$ magnet with
a magnetic field $B_{x}$ 
applied transverse to the Ho$^{3+}$ 
Ising spin direction~\cite{Wu,Bitko,Brooke},
which is parallel to the c-axis of the body-centered 
tetragonal structure of LiHo$_x$Y$_{1-x}$F$_4$~\cite{Hansen}.
Crystal field effects give an Ising
ground state doublet, $\vert \Phi_0^\pm\rangle$,
and a first excited singlet,  $\vert \Phi_{\rm e}\rangle$,
at approximately 9 K above the ground doublet~\cite{Hansen}. 
For ${x=1}$, LiHoF$_4$ is a 
dipolar-coupled ferromagnet (FM) with  ${T_c=1.53}$ K ~\cite{Bitko,Chakraborty}. 
Random disorder is generated 
by replacing the magnetic Ho$^{3+}$ ions by non-magnetic Y$^{3+}$.
Quantum mechanical (spin flip) fluctuations are induced
by $B_{x}$ which
admixes $\vert \Phi_{\rm e}\rangle$ with $\vert\Phi_0^\pm\rangle$,
splitting the latter, 
hence producing an effective TFIM with
${\Gamma=\Gamma(B_{x})}$  (${\Gamma\propto B_{x}^2}$  for small $B_{x}$)~\cite{Chakraborty}.  

Two experimental puzzles pertaining to the effect of $B_{x}$ on the FM
($0.25 < x <1.0$) and the SG ($x<0.25$) phases of LiHo$_x$Y$_{1-x}$F$_4$
have long been known.
Firstly, in the FM regime,  while the
mean-field argument that $T_c(x) \propto x$ for the PM to FM transition is well satisfied 
for $0.25<x<1.0$~\cite{Brooke}, the rate at which
$T_c(B_{x})$ is depressed by $B_{x}$ becomes
progressively 
faster than  mean-field theory predicts as $x$ is reduced~\cite{Brooke-thesis}.
This implies that, compared with the energy scale 
for FM order set by ${T_c(B_{x}=0)}$, 
$B_{x}$ becomes ever more efficient at destroying
FM order the lower $x$ is~\cite{Brooke-thesis}.
Secondly, for ${B_{x}=0}$,  LiHo$_{0.167}$Y$_{0.833}$F$_4$  displays a 
conventional SG transition, 
with a nonlinear magnetic susceptibility $\chi_3$ diverging at 
$T_g$ as $\chi_3(T)\propto (T-T_g)^{-\gamma}$~\cite{Mydosh}.
However, $\chi_3(T)$ becomes less singular
as $B_{x}$ is increased from ${B_{x}=0}$, 
with no indication that a QPT between PM and 
SG
states occurs as $T\rightarrow 0$~\cite{Wu,Wu-thesis}.
It has recently been suggested that
for dipole-coupled Ho$^{3+}$ ions 
nonzero $B_{x}$ 
generates both longitudinal (along the Ising $\hat z$ direction)~\cite{Us}
and transverse~\cite{Schechter-PRL,Schechter-2}
random fields (RFs) that 
either renormalize the critical transverse field~\cite{Us,Schechter-PRL},
or even destroy the SG transition~\cite{Schechter-2}.
In this paper we examine the quantitative merit
of this hypothesis 
by comparing results from numerical and analytical calculations 
with experimental results on LiHo$_x$Y$_{1-x}$F$_4$. We obtain 
compelling evidence that RFs are manifestly
at play in LiHo$_x$Y$_{1-x}$F$_4$ 
and explain the above two long-standing paradoxes.

We first show that
the low-energy effective theory of LiHo$_x$Y$_{1-x}$F$_4$
for $x<1$ and $B_{x}>0$ is 
a TFIM with additional $B_{x}-$induced RFs.
We start with 
Hamiltonian 
${{\cal H}={\cal H}_0+{\cal H}_{\rm dip}}$
expressed in 
terms of the angular momentum operator ${\bm J}$ of Ho$^{3+}$
(${J=8,L=6,S=2}$).
The single ion part, ${\cal H}_0=\sum_i [{\cal H}_{\rm cf}({\bm J}_i)
+{\cal H}_{\rm Z}({\bm J}_i)]$, 
consists
of the crystal field Hamiltonian,
${\cal H}_{\rm cf}({\bm J}_i)$, of Ho$^{3+}$ 
in the LiHo$_x$Y$_{1-x}$F$_4$ environment~\cite{Hansen,Chakraborty},
and the Zeeman field term, 
${\cal H}_{\rm Z}=-g{\mu_{\rm B}} ({\bm J}_i\cdot {\bm B})$,
with ${\bm B}$ the magnetic field.
$g=5/4$ is the Ho$^{3+}$ Land\'e factor and $\mu_{\rm B}$  is the Bohr magneton.
The interactions between ions are dominated by long-range magnetic dipolar 
interactions ${\cal H}_{\rm dip}$ \cite{Bitko,Chakraborty,exchange},  
$
{\cal H}_{\rm dip}=
({g^2\mu_{\rm B}^2}/{2})\sum_{(i,j)} {\epsilon_i\epsilon_j}
\left [
{\bm J}_i\cdot {\bm J}_j - 3
({{\bm J}_i\cdot {\bm r}_{ij} {\bm J}_j\cdot {\bm r}_{ij}})
{r_{ij}^{-2}}
\right ]
{r_{ij}}^{-3}
$,
where ${\bm r}_i$ are the crystalline positions occupied either by a magnetic Ho$^{3+}$ 
ion ($\epsilon_i=1$) or a non-magnetic Y$^{3+}$ ion 
($\epsilon_i=0$), and $r_{ij}=\vert {\bm r}_{j}-{\bm r}_{i}\vert$ is 
the inter-ion distance.

The two lowest energy eigenstates 
$\vert \Psi_i^+(B_{x})\rangle$ and
$\vert \Psi_i^-(B_{x})\rangle$ of ${\cal H}_0$ are sufficiently below 
$\vert\Phi_{\rm e}\rangle$
that the latter can be ignored at temperatures near and below 
${T_c(x=1)\sim 1.5}$ K~\cite{Chakraborty}.
This allows us to recast ${\cal H}$ in terms of an effective Hamiltonian,
${\cal H}_{\rm eff}$, with $S=1/2$ pseudo 
spin operators that act in the restricted 
low-energy subspace spanned by the 
$\prod_i \vert \Psi_i^{\sigma_i}(B_{x})\rangle$ ($\sigma_i=\pm$)
 eigenstates of ${\cal H}_0$.
In this subspace, 
$\Gamma(B_x) \equiv (1/2)
[\langle \Psi^+\vert {\cal H}_0 \vert \Psi^+\rangle -
 \langle \Psi^-\vert {\cal H}_0 \vert \Psi^-\rangle
]$. 
The projected $J_i^\mu$ ($\mu=x,y,z$) operator is written as:
$J_i^\mu=\sum_\nu C_{\mu\nu}(B_{x})\sigma_i^\nu +C_{\mu 0 }{\openone}$~\cite{Chakraborty}.
The $\vert +\rangle$ and
$\vert  - \rangle$ eigenstates of $\sigma_i^z$ are written in terms of 
$\vert \Psi_i^\pm\rangle$ 
such that $J_i^z=C_{zz}\sigma_i^z$~\cite{Chakraborty}. 
 For $B_{x}=0$, only $C_{zz}\ne 0$ and 
decreases slightly with increasing $B_{x}$, 
 while the other $C_{\mu \nu}(B_{x})$  parameters and 
  $\Gamma(B_{x})$  increase with $B_{x}$,
 starting from zero at $B_{x}=0$. 
By straightforward manipulations replacing $J_i^\mu$ in terms 
of $\sum_{\nu} C_{\mu\nu}\sigma_i^\nu$ 
in ${\cal H}_{\rm dip}$, one finds that the terms with largest 
$C_{\mu\nu}(B_x)$ in ${\cal H}_{\rm eff}$, including the
transverse field term $\Gamma\sigma_i^x$, are:
\begin{eqnarray}
\label{Heff}
  {\cal H}_{\rm eff}& =&  
  \frac{(g \mu_{\rm B})^2}{2} 
{C_{zz}}^2\sum_{(i,j)}\epsilon_i\epsilon_j L_{ij}^{zz}\sigma_i^z \sigma_j^z 
- \Gamma \sum_i \sigma_i^x + \\
& &\hspace{-12mm}  
{(g \mu_{\rm B})^2}
C_{zz}
\{C_{x0}\sum_{(i,j)} \epsilon_i \epsilon_j L_{ij}^{xz}  \sigma_i^z + 
C_{xx}\sum_{(i,j)} \epsilon_i \epsilon_j L_{ij}^{xz} \sigma_i^x \sigma_j^z
\}
\nonumber
\end{eqnarray}
where 
$L_{ij}^{ab}=[\delta_{ab}-3{r_{ij}^a} {r_{ij}^b}/{r_{ij}}^2]{r_{ij}}^{-3}$.
We see from Eq.~\ref{Heff}
that a longitudinal random field (RF) term $\propto \sigma_i^z$
along $\hat z$   and an
off-diagonal (bilinear) $\propto \sigma_i^x \sigma_j^z$ coupling
are induced by $B_{x}\ne 0$
($C_{x0}=C_{xx}=0$ for $B_{x}=0$). 
For pure LiHoF$_4$ ($x=1$, all $\epsilon_i=1$),  
lattice symmetries
enforces $\sum_j L_{ij}^{xz}=0$, causing
the term linear in $\sigma_i^z$ and the 
Boltzmann thermal average $\langle \sigma_i^x \sigma_j^z\rangle$ to vanish for $B_{x}>0$.
However, for ${x<1}$ and ${B_{x}>0}$, a RF $\propto \sigma_i^z$ emerges.
With the time-reversal symmetry broken by $B_{x}$ ($\langle J_i^x\rangle>0$), the 
bilinear $\sigma_i^x \sigma_j^z$ also provides a ``mean-field''
contribution to the longitudinal RFs, $\langle \sigma_i^x\rangle \sigma_j^z$, as well as 
transverse RFs,  $\langle \sigma_i^z\rangle \sigma_j^x$.
We find that $C_{xx}/C_{x0}\lesssim 1$ for all $B_{x}>0$ so that the leading correction
to $H_{\rm TFIM}$  is indeed a (correlated) RF term,
$\sim \sum_i h_i^z \sigma_i^z$,
with $h_i^z \propto C_{x0}C_{zz}\sum_j \epsilon_j L_{ij}^{xz}$.

We  now investigate the effect of the RFs on the PM to FM transition.
To do so, we make a mean-field (MF) approximation to ${\cal H}$ and consider the 
one-particle MF Hamiltonian, ${\cal H}_i^{\rm MF}$, for an arbitrary
site ${\bm r}_i$ occupied
 by a Ho$^{3+}$ moment:
${\cal H}_i^{\rm MF}={\cal H}_{\rm cf}-g\mu_{\rm B} ({\bm h}_i^{\rm MF}\cdot {\bm J}_i)
-g\mu_{\rm B} J_i^xB_{x}$. 
${\bm h}_i^{\rm MF}$ is the MF acting on magnetic moment ${\bm J}_i$,
${\bm h}_i^{\rm MF}=
g\mu_{\rm B} \sum_j 
\epsilon_j [3 ({\bm r}_{ij}\cdot {\bm M}_j){\bm r}_{ij}/r_{ij}^2- {\bm M}_j]{r_{ij}}^{-3}$,
with the self-consistent MF equation 
${\bm M}_i=\langle {\bm J}_i\rangle$
with
${\bm M}_i={\rm Tr} (\rho_{\rm MF} {\bm J}_i)/{\rm Tr} (\rho_{\rm MF})$ and
$\rho_{\rm MF}=\exp(-\beta H_i^{\rm MF})$
with $\beta\equiv 1/(k_{\rm B}T)$.
The crystal field Hamiltonian $H_{\rm cf}$ 
for Ho$^{3+}$ expressed in 
terms of the components  $J_i^\mu$ is taken from 
the 4 crystal field parameter (4CFP) ${\cal H}_{\rm cf}$
of Ref.~[\onlinecite{Hansen}].
Slightly different choices of ${\cal H}_{\rm cf}$~\cite{Hansen,Chakraborty} 
do not qualitatively affect the results.
We diagonalize ${\cal H}_i^{\rm MF}$ 
for each $i$ and update the ${\bm M}_i$'s 
using the newly obtained eigenstates. We continue iterating
until convergence is reached at the n'th iteration, defined by the convergence
criterion $\sum_i ({\bm M}_i^{(n)}-{\bm M}_i^{(n-1)})^2/
\sum_i ({\bm M}_i^{(n)})^2 \le 10^{-6}$.
To simplify the calculations and speed up the convergence we only
keep the diagonal 
$L_{ij}^{zz}$
and
the  off-diagonal $L_{ij}^{xz}$
terms in ${\cal H}_{\rm dip}$ since no
qualitatively new physics arises when keeping the other 
diagonal ($L_{ij}^{xx}$ and $L_{ij}^{yy}$)
 and off-diagonal ($L_{ij}^{xy}$, $L_{ij}^{yz}$) 
terms in ${\cal H}_{\rm dip}$. 
For a given $B_{x}$, we compute the
temperature dependence of the magnetization, $M_z=[\sum_i M_i^z]_{{\rm d}}$,
 along the $\hat z$ direction~\cite{finite-size}.
Here $[\ldots]_{\rm d}$ signifies an average over the bimodal
lattice occupancy probability distribution,
$P(\epsilon_i)=x\delta(\epsilon_i-1)+(1-x)\delta(\epsilon_i)$.
The transition temperature $T_c(B_{x})$
 is determined by the temperature 
at which $M_z(T)$ sharply rises~\cite{finite-size}.
We consider a system of linear size $L=4$, with
total number of sites $N_0=4L^3$ with $N=xN_0$ sites occupied by Ho$^{3+}$, 
and perform disorder averages over 50 different diluted samples.
 We use the
Ewald summation method to define 
infinite-range dipolar interactions. 
For $B_{x}=0$
we find that, as found experimentally, 
the decrease of $T_c$ while reducing $x$ is proportional to $x$~\cite{Brooke,Brooke-thesis}.
To investigate the role of RFs, we compare $T_c(B_{x})$ obtained 
when both the $L_{ij}^{xz}$ and $L_{ij}^{zz}$
 terms in ${\cal H}_{\rm dip}$ are kept (open symbols)
with the $T_c(B_{x})$ found when only the $L_{ij}^{zz}$ term is retained (filled symbols).
\begin{figure}[t]
\includegraphics[angle=0,height=5cm,width=8cm]{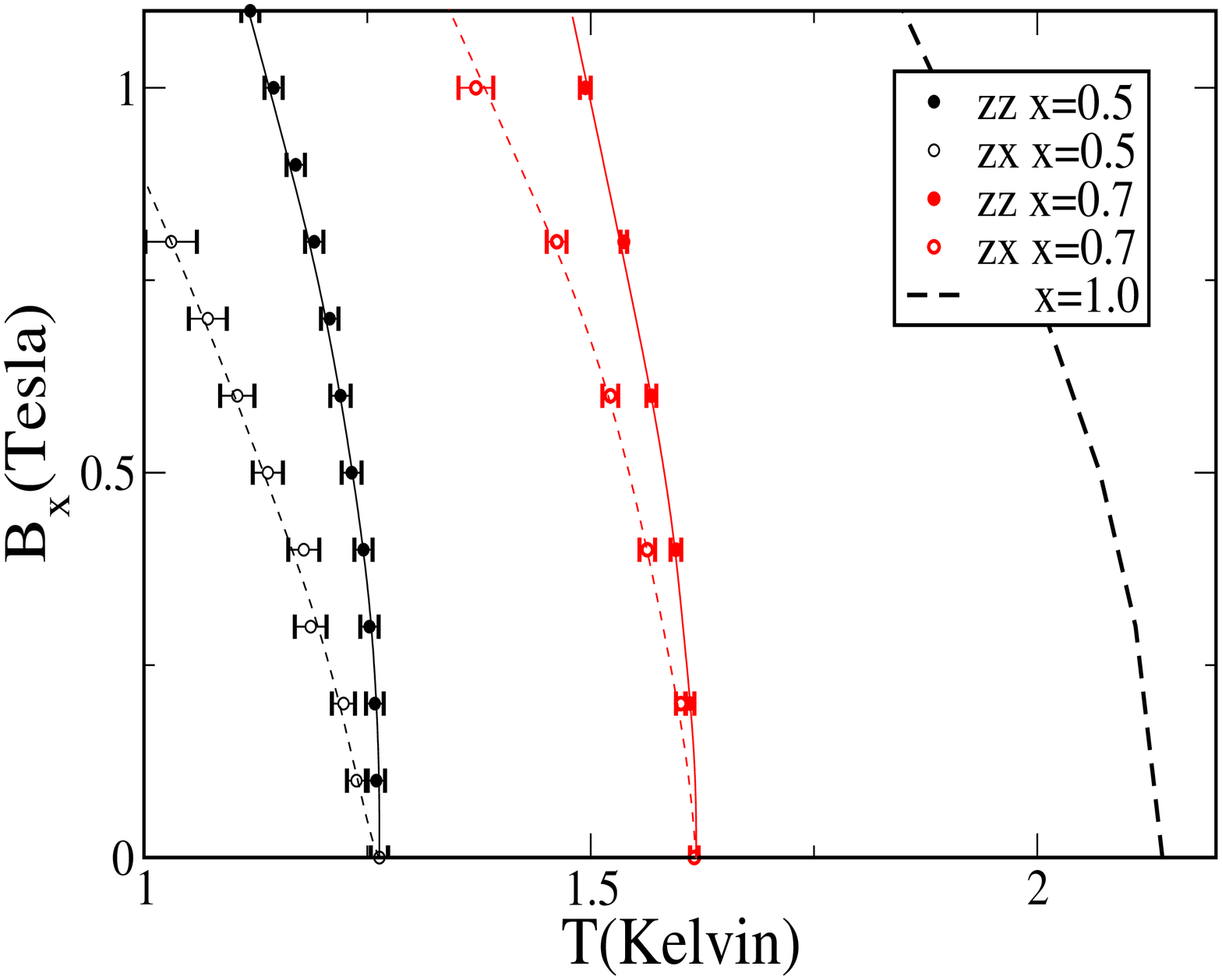}
\caption{
$T_c(B_{x})$ from a finite lattice mean-field calculation (see text).
For $x=0.7$ and $x=0.5$, the filled symbols show $T_c(B_{x})$ when
only the $L_{ij}^{zz}$
dipolar terms that couple $z$ compoments of ${\bm J}_i$ are kept.
The open symbols show the increased rate of depression of $T_c(B_{x})$ when
the off-diagonal dipolar $L_{ij}^{xz}$ couplings are 
included.
For the pure $x=1$ case, $T_c(B_{x})$ is the same for both models since
the lattice symmetries eliminate the internal local random fields along 
$\hat z$.
}
\end{figure}
Figure 1 shows  that $T_c(B_{x})$ is depressed for small $B_{x}$ 
faster when the off-diagonal $L_{ij}^{xz}$ term in  ${\cal H}_{\rm dip}$ is 
present than when only the Ising $L_{ij}^{zz}$ term is kept.
Also, as found experimentally~\cite{Brooke-thesis}, 
the rate at which $T_c(B_{x})$ is depressed 
by  $B_{x}$ increases as $x$ is lowered.
This provides strong evidence that the experimental observation
is due to $B_{x}-$induced RFs whose variance  increases as
$x$ decreases or $B_x$ increases.

We now consider 
the role of RFs at the SG transition.
While the nonlinear susceptibility $\chi_3$ diverges
at $T_g\simeq 0.13$ K in LiHo$_{0.167}$Y$_{0.833}$F$_4$  when $B_{x}=0$,
$\chi_3(B_{x},T)$ becomes progressively smeared as $B_{x}$ is turned on
(see top inset of Fig. 3)~\cite{Wu,Wu-thesis}.
It has been suggested that random off-diagonal dipolar couplings
destroy the SG transition
when $B_{x}>0$~\cite{Schechter-2}.
Here, we take a more pragmatic approach and
ask whether the 
behavior of 
$\chi_3(B_{x}>0,T)$ in SG samples of LiHo$_{x}$Y$_{1-x}$F$_4$ 
can indeed be interpreted in terms of induced RFs.
To investigate this question
on $\chi_3(B_{x},T)$,
we introduce a 
mean-field variant of ${\cal H}_{\rm eff}$ in Eq. (1) 
that preserves the crucial physics therein. Our model is a
generalization of the  
Sherrington-Kirkpatrick transverse field Ising SG model~\cite{Sen,Kopec,Guo,Kim}
but with additional off-diagonal and RF interactions similar to that in 
${\cal H}_{\rm eff}$:
\begin{eqnarray}
{\tilde{\cal H}}& = & \frac{1}{2}\sum_{(i,j)} J_{ij}\sigma_i^z \sigma_j^z 
       \;+ \; \frac{1}{2}\sum_{(i,j)}K_{ij}\sigma_i^x\sigma_j^z \; - \nonumber \\
& &  \Gamma\sum_i \sigma_i^x -\sum_i h_i^z \sigma_i^z -h_0^z\sum_i \sigma_i^z
\;\;\; .
\end{eqnarray}
For simplicity, we take infinite-ranged $J_{ij}$ and $K_{ij}$  
given by independent
Gaussian distributions of zero mean and variance $J^2$ and $K^2$, respectively.
$h_i^z$ is a Gaussian RF with zero mean and variance $\Delta^2$.
Model (2), but with $K_{ij}=0$,
 was previously used to calculate $\chi_3$ in 
quadrupolar glasses~\cite{Kopec},
which also possess internal RFs~\cite{Kopec,Holdsworth}. 

We follow the procedure  of Ref.~\cite{Kim},
employing the imaginary time formalism and the replica trick to derive
the (replicated) free energy  of the system. 
To further simplify the calculations, we make a static approximation for the
replica-symmetric solution in the PM phase.
This allows us to derive self-consistent equations 
for the $\alpha$ components of the magnetization, $M_\alpha$,
and spin-glass order parameters $Q_\alpha$ ($\alpha=x,z$).
We find
$
M_\alpha  = (1/{2\pi})\int_{-\infty}^{\infty} 
dxdz \,  {\rm e}^{-(x^2+z^2)/2}
\, [(H_\alpha/H)\tanh(\beta H)]^{p}
$
with $p=1$,
$H_z=(h_0^z+z\sqrt{J^2Q_z+(K^2/{2})Q_x+\Delta^2})$,
$H_x=(\Gamma + x\sqrt{(K^2/2)Q_z})$
and $H^2={H_x}^2+{H_z}^2$.
The self-consistent equations for $Q_x$ and $Q_z$ are obtained by replacing 
$M_\alpha\rightarrow Q_\alpha$ above and setting $p=2$. 
$\chi_3$ is obtained 
from $\chi_3=\frac{1}{6}\partial^3M_z/{( \partial  {h_0^z} )}^3$, 
and by solving numerically the resulting four coupled
self-consistent equations.
Figures (2a) and (2b) show the $\Gamma$ dependence of $\chi_3$ for various
temperatures, $T$,
in models either  with only $h_i^z$ random fields (Fig. (2a))
or with only random off-diagonal $K_{ij}$ couplings 
(Fig. (2b)).
\begin{figure}[t]
\begin{center}
\includegraphics[width=2.6in,height=1.6in]{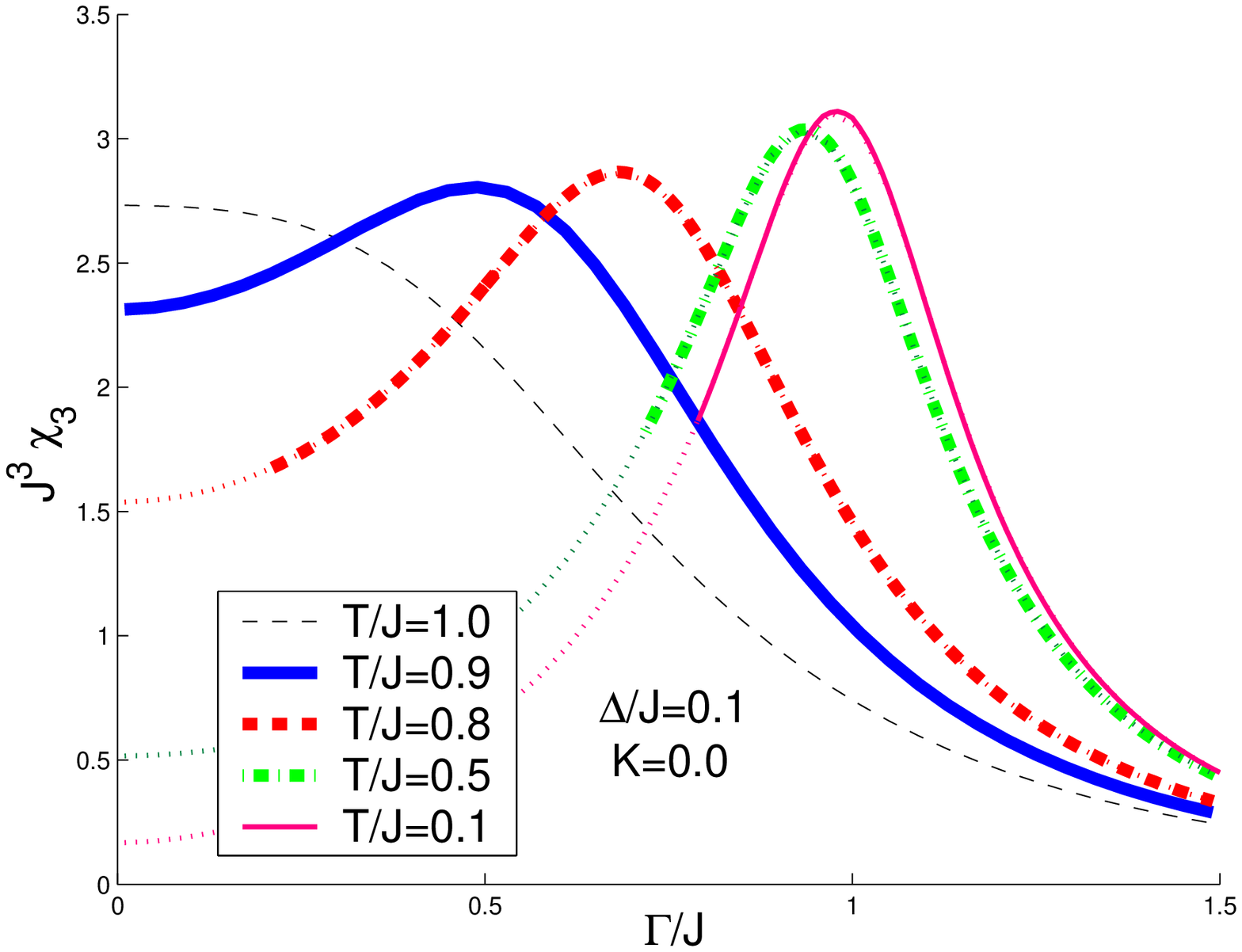}
\includegraphics[width=2.6in,height=1.6in]{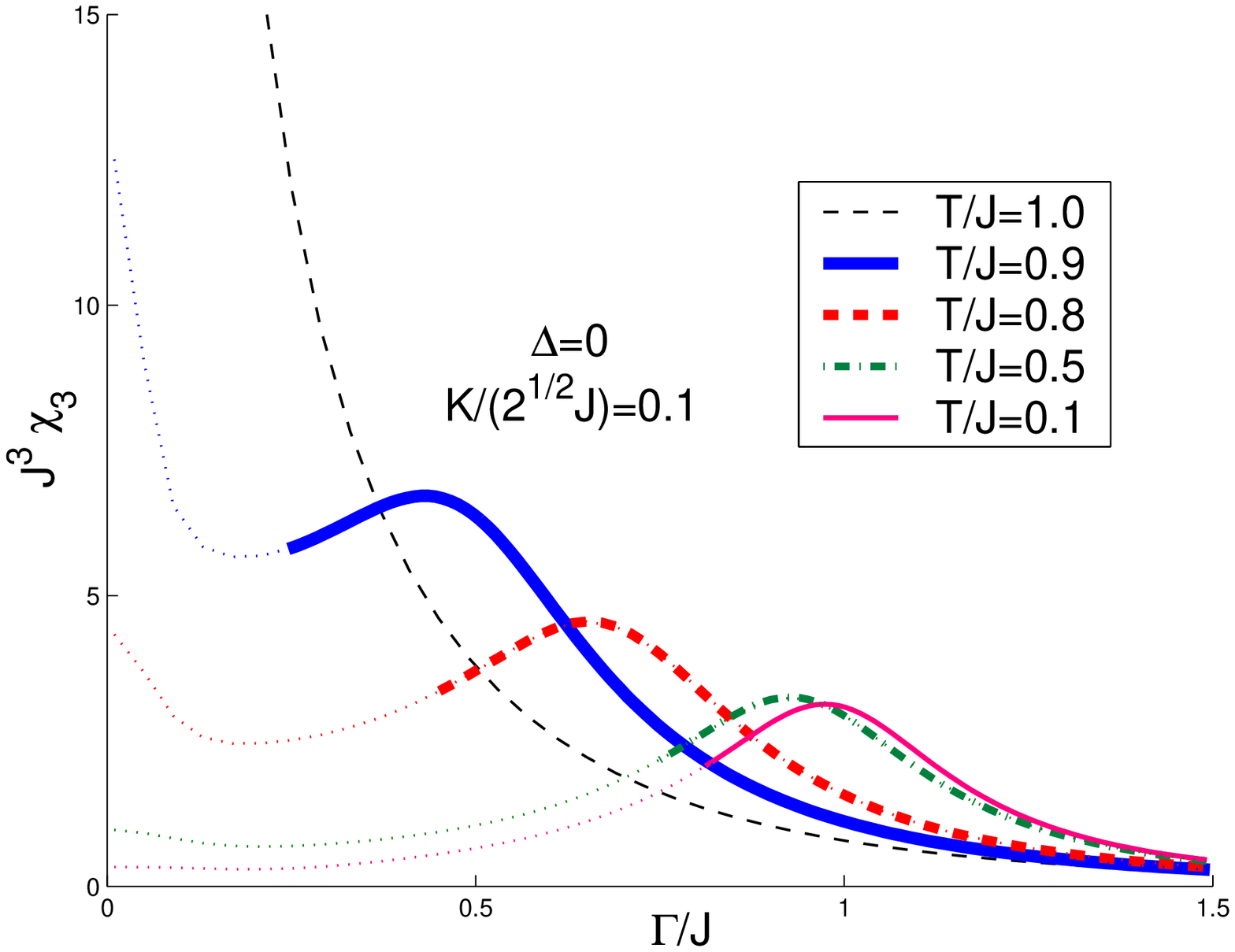}
\caption{(a), top panel: $\chi_3$ vs $\Gamma$ with $K=0$.
(b), bottom panel: $\chi_3$ vs $\Gamma$ with $\Delta=0$.
The change of linestyle at low $\Gamma$ indicates the limit of stability of
the replica symmetric paramagnetic solution, as determined following the
standard procedure~\cite{Kim}.
}
\label{fig2}
\end{center}
\end{figure}

Comparing Fig. (2) with the experimental $\chi_3^{\rm exp}(B_{x})$ (top inset of Fig. (3)),
one finds that the dependence of $\chi_3$ upon $\Gamma $ in Fig. (2a) does not
show a decrease in magnitude as $T$ is decreased. 
Also, while $\chi_3$ shows a decreasing amplitude with decreasing $T$ in Fig. (2b), 
it does not reveal a rapid sharpening 
as $T$ is increased.
It turns out that the key physics
ingredient missing in these calculations is the underlying {\it microscopic}
dependence of $J$, $\Delta$ and $K$ upon $B_{x}$ 
via the $C_{\mu\nu}(B_{x})$ transformation coefficients.
Physically, the built-in $B_{x}$ dependence of the
$C_{\mu\nu}$ insures that the $B_x$ scale
at which $\chi_3$ is quenched by RFs 
is not trivially tied to the scale at which 
the $T=0$ SG-PM crossover occurs, 
as it is in
${\tilde{\cal H}}$
with $J_{ij}$, $K_{ij}$, $\Gamma$ and $h_i^z$ 
independent of $B_{x}$ (c.f Fig. (2)).
The widths $J(B_{x})$, $\Delta(B_x)$ and $K(B_{x})$  are obtained by calculating
the disorder average of the first, third and fourth lattice sums in Eq. (1), respectively.
We have
$K^2$ $=$ 
$4(g\mu_{\rm B})^4 [C_{zz}(B_{x})C_{xx}(B_{x})]^2 [\frac{1}{N_0}\sum_{(i,j)}
\epsilon_i\epsilon_j ({L_{ij}^{xz}})^2]_{\rm d}$
and
$\Delta^2$ $=$ 
$(g\mu_{\rm B})^4 [C_{zz}(B_{x})C_{x0}(B_{x})]^2
[\frac{1}{N_0}\sum_{i}
\epsilon_i(\sum_{j\neq i}\epsilon_j {L_{ij}^{xz}})^2]_{\rm d}$.
To make further contact between calculations and
experimental data,
we note that hyperfine interactions,
which are important in Ho-based materials~\cite{Bitko,Chakraborty},
lead to a renormalization of the critical $B_x$, $B_x^c$,
when
$T_c$ or
$T_g$ is less than the hyperfine energy scale~\cite{Chakraborty,Schechter-PRL}.
To obtain a relation between $\Gamma$ and 
$B_{x}$ in  ${\tilde{\cal H}}$,
which does not incorporate hyperfine effects,
we set $\Gamma(B_{x})/{J(B_{x})}$ $=$ $1.05(B_{x}/B_{x}^c)^{0.35}$, where $B_{x}^c=1.2$ T
is the experimental zero temperature
critical field~\cite{Wu-thesis}.
This ansatz for $\Gamma(B_{x})$ is obtained by matching  
the critical temperature $T_g(B_{x})$  of LiHo$_x$Y$_{1-x}$F$_4$ 
with the  $T_g(\Gamma)$ of ${\tilde{\cal H}}$.
For the former we use 
$T_g(B_{x})$ $=$ $T_g(0)[ 1-\left(B_{x}/B_{x}^c\right)^\phi] $
($\phi \approx 1.7$) as found experimentally~\cite{Wu-thesis}.
For ${\tilde{\cal H}}$,
we find 
$T_g(\Gamma)\approx T_g(0)[ 1-a\left(\Gamma/T_g(0)\right)^\psi ]$
($a\approx 0.79$ and $\psi\approx 4.82$) by fitting $T_g(\Gamma)$ vs $\Gamma$ 
with $K=\Delta=0$.
To incorporate the role of hyperfine effects
 on $h_i^z$ and
$K_{ij}$, we rescale $K$ and $\Delta$ calculated from their microscopic
origin in Eq. (1) by a scale factor $\eta=0.15$ 
that positions the peak of $\chi_3^{\rm theo}(B_x)$
at $B_x\sim 5$ kOe for $T_g(B_x)/T_g(0)=0.75$.
With an estimate of $\eta$ available, one could 
then determine a parametric $\Gamma(B_{x})$ such that, for
a given $T/T_g(B_{x}=0)$, the
experimental $\chi_3^{\rm exp}$ and theoretical $\chi_3^{\rm theo}$ peak
at the same $B_{x}$ value for all $T_g(B_x)/T_g(0)$.
Such procedure gives results qualitatively very 
similar to those in Fig. 3.

\begin{figure}[t]
\includegraphics[angle=0,width=8cm]{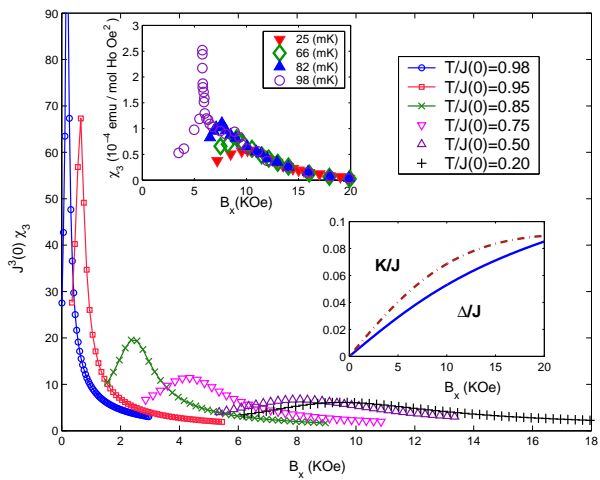}
\caption{
$\chi_3^{\rm theo}$ vs $B_{x}$ for different temperatures given by the model Eq. (2).
The top inset shows $\chi_3^{\rm exp}$ from Ref.~[\onlinecite{Wu}].
The parameters $K$ and $\Delta$ are computed using
$x=0.167$
and rescaled by $\eta=0.15$
are shown in the bottom inset (see text).
}
\end{figure}

 $\chi_3^{\rm theo}$ 
reproduces the overall
trend of 
 $\chi_3^{\rm exp}$ 
This is evidence
that RFs are indeed at play when $B_{x}>0$ and at the origin of the 
experimental $\chi_3(B_{x},T)$ behaviors in 
SG samples of
LiHo$_x$Y$_{1-x}$F$_4$~\cite{Wu,Wu-thesis}.
A noticeable difference between 
$\chi_3^{\rm theo}$ and $\chi_3^{\rm exp}$,  
is that,
for fixed $T/T_g(0)$, $\chi_3^{\rm theo}$ collapses
more rapidly and at smaller $B_{x}$ than 
$\chi_3^{\rm exp}$. 
This is likely further manifestation of
the renormalization effects 
of the RFs 
and random off-diagonal couplings
caused by the aforementioned hyperfine interactions.
We will report elsewhere results addressing this hyperfine renormalization
of ${\cal H}_{\rm eff}$ and its role on $\chi_3(B_{x},T)$.

In conclusion, by comparing numerical and analytical results with
experimental data, we have obtained compelling evidence that 
induced random fields (RFs) are indeed at play and ``observed'' in
LiHo$_x$Y$_{1-x}$F$_4$. 
As a result,
LiHo$_x$Y$_{1-x}$F$_4$ in a transverse field (TF) is identified as a new  RF
Ising system.
As found in other RF systems, we expect
that the nontrivial
fixed point of the theory (${\cal H}_{\rm eff}$)
is controlled by a fluctuationless {\it classical}
zero temperature fixed point.
In particular, this is what
occurs
in a TFIM plus RFs (${\tilde{\cal H}}$ with $K_{ij}=0$)~\cite{Senthil}.
Hence, it would therefore seem that
quantum criticality is most likely innaccessible
in ferromagnetic LiHo$_x$Y$_{1-x}$F$_4$  samples.
For the Ising spin glass (SG) model with RFs
along
$\hat z$,
recent studies suggest that 
there is 
no Almeida-Thouless line~\cite{Mydosh}
 and no thermodynamic SG transition~\cite{Katzgraber}.
Hence, from the arguments above leading to ${\cal H}_{\rm eff}$,
$B_{x}-$induced
SG to PM quantum criticality in LiHo$_x$Y$_{1-x}$F$_4$  
would also appear likely inexistent.
The presence of $B_{x}-$induced RFs and the quenching of quantum criticality
is presumably the reason why, unlike in quantum Monte Carlo simulations 
of TF Ising SG models~\cite{Guo},
Griffiths-McCoy singularities~\cite{GMS} have not been reported 
in LiHo$_x$Y$_{1-x}$F$_4$~\cite{Wu,Brooke}. 
While quantum criticality seems unlikely for {\it any} 
$B_{x}>0$ and $x<1$,
a new set of interesting questions has nevertheless 
arisen: do 
 FM samples of LiHo$_x$Y$_{1-x}$F$_4$ 
 with $B_{x}>0$ indeed display 
classical RF criticality and all the fascinating phenomena of the 
RF Ising model~\cite{Belanger}?
Given the scarcity
of real RF Ising materials~\cite{Belanger}, the identification
of another such system opens new avenues for future theoretical and experimental
investigations.



We thank D. Belanger, S. Girvin, B. Malkin, M. Moore,
T. Rosenbaum, T. Yavors'kii and A.P. Young
for useful discussions. 
This work was supported by the NSERC of Canada, 
the Canada Research Chair Program (Tier I, M.G),
Research Corporation and the CIAR.
M.G. thanks the U. of Canterbury (UC) for an Erskine Fellowship
and the hospitality of
the Department of Physics and Astronomy at UC 
where part of this work was done.
Y.K. is partially supported by the NSC of Taiwan


\begin{thebibliography}{99}
\bibitem{exchange}
We ignore a small nearest-neighbor exchange interaction in
LiHo$_x$Y$_{1-x}$F$_4$
 which does not play an important role
in the physics that we are discussing here (see Ref. [10]).
\bibitem{finite-size}
 Because of finite-size effects, there is a residual net field along the
$\pm \hat z$ direction for any given realization of disorder that dictates the
direction of $M_z$ at low temperatures.
\end{thebibliography}

\end{document}